\documentclass[prb,twocolumn,
superscriptaddress,showpacs,
secnumarabic]{revtex4}
\usepackage{amsmath,amsfonts,amssymb,graphicx
}

\newcommand{\Eq}[1]{Eq.~\eqref{#1}}
\newcommand{\eq}[1]{\eqref{#1}}
\newcommand{\Fig}[1]{Fig.~\ref{#1}}

\newcommand{\beq}{\begin{equation}}
\newcommand{\eeq}{\end{equation}}
\newcommand{\beqa}{\begin{eqnarray}}
\newcommand{\eeqa}{\end{eqnarray}}
\newcommand{\Beqa}{\begin{eqnarray*}}
\newcommand{\Eeqa}{\end{eqnarray*}}
\newcommand{\nn}{\nonumber}

\newcommand{\pdag}{{\phantom{\dagger}}}
\newcommand{\pprime}{{\phantom{\prime}}}

\begin{document}

\title{
Thermalization of acoustic excitations \\ in a strongly interacting one-dimensional quantum liquid
}
\author{Jie Lin}
\affiliation{Materials Science Division, Argonne National Laboratory, Argonne, IL 60439}
\author{K. A. Matveev}
\affiliation{Materials Science Division, Argonne National Laboratory, Argonne, IL 60439}
\author{M. Pustilnik}
\affiliation{School of Physics, Georgia Institute of Technology, Atlanta, GA 30332}

\begin{abstract}
We study inelastic decay of bosonic excitations in a Luttinger liquid. In a model with linear excitation spectrum the decay rate diverges. We show that this difficulty is resolved when the interaction between constituent particles is strong, and the excitation spectrum is nonlinear.  Although at low energies the nonlinearity is weak, it regularizes the divergence in the decay rate. We develop a theoretical description of the approach of the system to thermal equilibrium. The typical relaxation rate scales as the fifth power of temperature.
\end{abstract}


\date{August 16, 2012}

\maketitle

One-dimensional interacting systems~[\onlinecite{Giamarchi}] are fundamentally different from their higher-dimensional counterparts~[\onlinecite{Pines}]. Regardless of the statistics of the constituent particles, elementary excitations in one dimension are believed to be bosons~[\onlinecite{Haldane,LL_bosons,Giamarchi}], the waves of density. Similar to sound waves in ordinary fluids, bosonic excitations in such a \textit{Luttinger liquid}~[\onlinecite{Haldane,LL_bosons}] have linear spectrum at low energies $\omega_q = s|q|$. Here $q$ is the wave number and $s$ is the velocity.

Just as quasiparticles in the Fermi liquid~[\onlinecite{Pines}], bosons in the Luttinger liquid do not represent exact eigenstates of a generic one-dimensional system. At finite energies, the corresponding effective Hamiltonian should be amended by irrelevant in the renormalization group sense perturbations~[\onlinecite{Haldane}], such as interaction between the bosons. However, a naive attempt to account for this interaction perturbatively immediately leads to difficulties. 

\begin{figure}[b]
\centering
\includegraphics[width=.99\columnwidth]{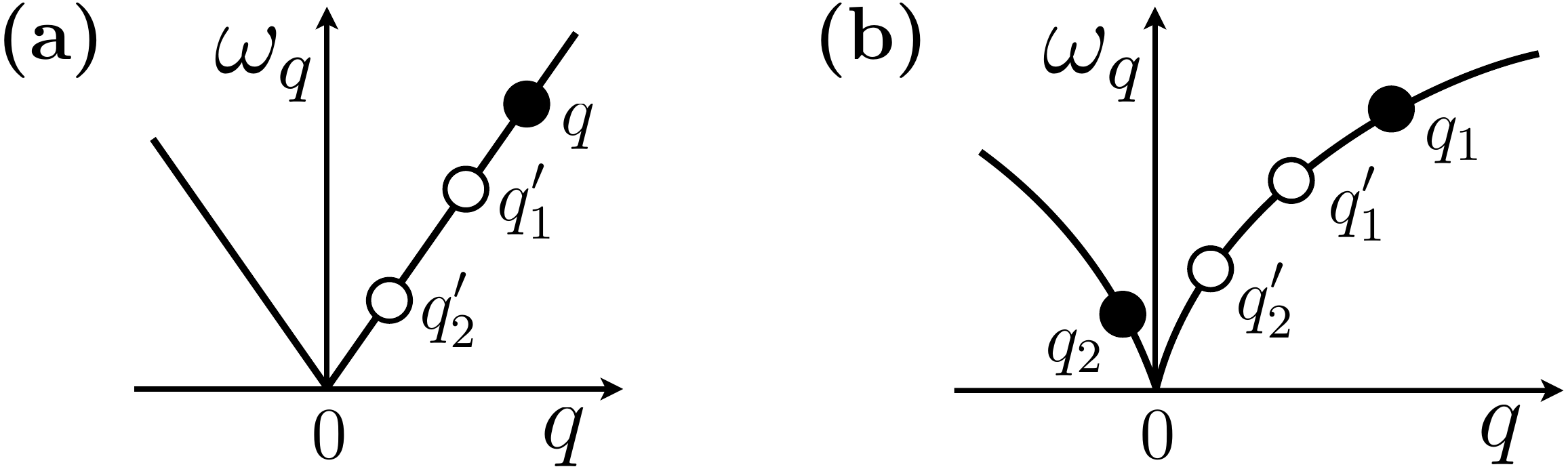}
\caption{
(a) For bosons with a linear spectrum scattering of a single boson (filled circle) into two (open circles) has a divergent rate.
(b) For bosons with a nonlinear spectrum the simplest scattering event satisfying the momentum and energy conservation laws involves two bosons both in the initial state (filled circles) and in the final state (open circles). For given $q_1$ and $q_2$, the conservation laws yield a unique set $q_1^\prime,q_2^\prime$, thus leading to a finite transition rate.
}
\label{Fig1}
\end{figure}

Consider, for example, the interaction-induced decay of a boson with wave number $q$ into two bosons with wave numbers $q_1^\prime$ and $q_2^\prime$, 
see Fig.~\ref{Fig1}(a). The corresponding inelastic scattering rate is given by the Fermi golden rule, 
\beq
\tau_q^{-1}\!\propto 
\!\int\!dq_1^\prime dq_2^\prime \,[\ldots]\,\delta(q-q_1^\prime-q_2^\prime)\,\delta\bigl(\omega_q-\omega_{q_1^\prime}-\omega_{q_2^\prime}\bigr),
\label{1}
\eeq
where the two $\delta$-functions express the momentum and energy conservation. 
When all three wave numbers have the same sign, the second $\delta$-function reduces to $s^{-1}\delta(q-q_1^\prime-q_2^\prime)$, and the rate \eq{1} diverges.

One way around the failure of the perturbation theory is to abandon the effective Luttinger liquid description altogether and approach the problem from the original fermionic perspective~[\onlinecite{drag,PKKG}]. Indeed, for noninteracting fermions the spectral weight of the dynamic structure factor (Fourier transform of the density-density correlation function)
at a fixed $q$ is spread uniformly over a narrow interval of the width
\beq
\delta\omega_q = \hbar \rho^2 q^2\!/m_*
\label{2}
\eeq 
about $\omega = \omega_q$. Here $m_*$ is the effective mass, which for free fermions coincides with the bare mass $m$, and $q$ is the dimensionless (measured in units of the particle density $\rho$) wave number. At sufficiently small $q$, \Eq{2} is applicable to interacting fermions as well~[\onlinecite{drag,PKKG,Pereira}]. The inverse of the width,
$1/\delta\omega_q$, provides a natural estimate of the lifetime of bosons in the Luttinger liquid. Since $\delta\omega_q\propto\omega_q^2$, the bosons indeed represent well-defined quasiparticles. 

In this Letter we develop an alternative approach, based on the observation that
divergences that plague the evaluation of the quasiparticle decay rate in the conventional Luttinger liquid theory can be cured if the boson spectrum is nonlinear, such as
\beq
\omega_q = s|q|\bigl(1 - \xi q^2\bigr). 
\label{e3}
\eeq
Even for a weak nonlinearity $\xi q^2\ll 1$, decay of a single boson into two is forbidden by the momentum and energy conservation laws and can only occur virtually. The simplest \textit{real} scattering process involves two bosons both in the initial and in the final states, see \Fig{Fig1}(b), and has a finite rate.

Keeping the nonlinear correction in \Eq{e3} is justified only in the limit of strong repulsion, i.e., when the Luttinger liquid parameter~[\onlinecite{Giamarchi}] $K = \pi\hbar\rho^2/ms$ is small. Indeed, the correction must exceed the width $\delta\omega_q$ [see \Eq{2}], which can be viewed as an uncertainty in the energy of the Luttinger liquid's boson. Using the estimate~[\onlinecite{mass}] $m_*/m\sim\sqrt{K}$, valid for $K\ll 1$, we arrive at the condition $\xi q\gg \sqrt{K}$. 

For $K\ll 1$, \Eq{e3} is applicable in a broad range of wave numbers $\sqrt{K}\ll \xi q\ll \sqrt{\xi}$, and spectrum nonlinearity has a dramatic effect on inelastic scattering. For the scattering process with two bosons ($q_1$ and $q_2$) in the initial state and two bosons ($q_1^\prime$ and $q_2^\prime$) in the final state [see \Fig{Fig1}(b)], the conservation laws $q_1 + q_2 = q_1^\prime + q_2^\prime$ and 
$\omega_{q_1} + \omega_{q_2} = \omega_{q_1^\prime} + \omega_{q_2^\prime}$
yield a unique set $q_1^\prime,q_2^\prime$ for given $q_1,q_2$. Moreover, if $q_1^\pprime, q_1^\prime$, and $q_2^\prime$ belong to the same (say, right-moving) branch of the spectrum [see \Fig{Fig1}(b)], the remaining wave number is given by 
$q_2^\pprime \approx - \,(3\xi/2)q_1^\pprime q_1^\prime q_2^\prime$, i.e.,
the sign of $q_2$ is \textit{opposite} to that of $q_1^\pprime,q_1^\prime,q_2^\prime$ and the momentum transferred from the left-moving branch of the spectrum in each act of scattering is parametrically small compared with that redistributed among the three right-moving bosons. Accordingly, the process resembles decay of a single right-moving boson into two. However, unlike for bosons with strictly linear spectrum, the mere presence of the left-moving boson with a very small momentum, as required by the conservation laws, is sufficient to regularize the divergences.  

We describe our strongly interacting system by the Hamiltonian 
\beq
H=\sum_l\frac{\,p_l^2}{2m\,}+\frac{1}{2}\sum_{l\neq l^\prime} V(x_l-x_{l^\prime}),
\label{e4}
\eeq
where $p_l$ and $x_l$ are, respectively, the momentum and position of the $l$th particle $(l = 1,\ldots,N)$, and $V(x)$ is the interaction potential. In the strong repulsion limit (i.e., for $d^2 V/dx^2\bigr|_{x=1/\rho} \gg \hbar^2\rho^4\!/m$, which is equivalent to $K\ll 1$) the particles, regardless of their statistics, form at low energies a periodic chain, the so-called \textit{Wigner crystal} (see~[\onlinecite{Meyer-09}] for a review).
Although in one dimension quantum fluctuations destroy the true long-range order~[\onlinecite{Schulz-93}], the interparticle distance remains close to $1/\rho$. 

Similar to ordinary crystals, the elementary excitations of the Wigner crystal are \textit{phonons}. These phonons are nothing but the waves of density, with a typical for phonons linear dispersion at small momenta, i.e., the phonons coincide with the bosons of the effective Luttinger liquid theory. The boson spectrum $\omega_q$ in the leading (zero) order in $\hbar$ can be found by expanding the potential energy in \Eq{e4} to second order in the displacements of the particles from the corresponding lattice sites $u_l = x_l - l/\rho$, and solving classical equations of motion~[\onlinecite{MAP}]. For small $q$, this yields \Eq{e3} with model-dependent $s$ and $\xi$~[\onlinecite{suppl}]. 
 
Interaction between the bosons arises from the higher-order (anharmonic) terms in the expansion of the potential energy in \Eq{e3} in the displacements $u_l$. A scattering process with two bosons both in the initial and in the final states, see \Fig{Fig1}(b), can occur either in the first order in the quartic anharmonicity, or in the second order in the cubic anharmonicity~[\onlinecite{MAP}], and the corresponding contributions to the on-shell scattering amplitude $t_{q_1^\pprime q_2^\pprime;q_1^\prime q_2^\prime}$ are of the same order of magnitude. If all four wave numbers are small, the amplitude simplifies~[\onlinecite{suppl}] to
\beq
t_{q_1^\pprime q_2^\pprime;q_1^\prime q_2^\prime}
= \frac{\lambda}{N} \frac{\hbar^2 \rho^2}{m\,}\bigl|q_1^\pprime q_2^\pprime q_1^\prime q_2^\prime\bigr|^{1/2}
\,.
\label{e5}
\eeq
This expression is easy to understand if one notices that each boson with wave number $q$ participating in scattering contributes a factor of $(\hbar/\omega_q)^{1/2}|q|\propto (\hbar |q|)^{1/2}$ to the amplitude. The dimensionless parameter $\lambda$ in \Eq{e5} depends on the functional form of $V(x)$~[\onlinecite{suppl}]. In particular, $\lambda = 0$ for $V(x)\propto 1/\sinh^2(c\rho x)$ and $V(x) \propto 1/x^2$~[\onlinecite{suppl}], as expected for integrable models~[\onlinecite{Sutherland}] exhibiting no relaxation. For a generic interaction potential $|\lambda|$ is of order unity. In particular, $\lambda = -3/4$ for screened Coulomb interaction, see~[\onlinecite{suppl}].

Inelastic scattering leads to the relaxation of the boson distribution function $N_q$ towards equilibrium. The evolution of $N_q$ is described by the Boltzmann equation, which for a homogeneous system in the absence of external fields has the form~[\onlinecite{Lifshitz}]
\beq
\frac{\partial N_{q}}{\partial t} 
=
\mathcal{I}_\text{out}\bigl[N_{q}\bigr] 
+ \mathcal{I}_\text{in}\bigl[N_{q}\bigr],
\label{e6}
\eeq
where the two terms in the right-hand side describe, respectively, the scattering out of single-boson state $q$, and the scattering into this state. In the leading order in $\hbar$, these terms are given by
\Beqa
\mathcal{I}_\text{out}\bigl[N_{q}\bigr] 
&=&  -\sum_{p}\!\sum_{q_1>q_2}\!W_{q,p;q_1, q_2}
N_{q}N_{p}(1+N_{q_1})(1+N_{q_2}),
\\
\mathcal{I}_\text{in}\bigl[N_{q}\bigr] 
&=&  \sum_{p}\!\sum_{q_1>q_2}\!
W_{q,p;q_1,q_2}
(1+N_{q})(1+N_{p})N_{q_1}N_{q_2}
\Eeqa
with
\beqa
W_{q_1^\pprime, q_2^\pprime;q_1^\prime, q_2^\prime}
&=& \frac{2\pi}{\hbar^2}\,\bigl|t_{q_1^\pprime q_2^\pprime;q_1^\prime q_2^\prime}\bigr|^2
\delta_{q_1^\pprime + q_2^\pprime,q_1^\prime + q_2^\prime}
\nn \\
&&
\times\,\delta(\omega_{q_1}+\omega_{q_2}-\omega_{q_1^\prime}-\omega_{q_2^\prime}).
\label{e7}
\eeqa

We begin the analysis of Eqs. \eq{e6}-\eq{e7} by considering the relaxation rate of a single high-energy boson. Specifically, we assume that the distribution function $N_q$ differs from its equilibrium form, the Bose distribution $n_q = \bigl(e^{\hbar\omega_q/T} - 1\bigr)^{-1}$,  in the population of a single state with $q$ in the range $T/\hbar s\ll q\ll 1/\sqrt{\xi}$. In this limit $\mathcal{I}_\text{in}\bigl[N_{q}\bigr]$ is exponentially suppressed, and \Eq{e6} reduces to $\partial N_q/\partial t = - N_q/\tau_q$ with the relaxation rate
\beq
\tau_q^{-1}
= 
\frac{\lambda^2 K^2 s}{\,48\pi^3}\!
\times\!
\left\lbrace
\begin{array}{lc}
(T/\hbar s) q^4,
&
q\ll (T/\hbar s \xi)^{1/3}
\\
\\
\dfrac{a(T/\hbar s)^3}{(\xi q)^2}\,,
&
q\gg (T/\hbar s \xi)^{1/3}
\end{array} 
\right.,
\label{e8}
\eeq
where $a = 32\zeta(3)/3$. Here $\zeta(x)$ is the Riemann's zeta-function, $\zeta(3)\approx 1.2$. 
Although \Eq{e8} is not directly applicable to \textit{thermal} bosons with energy of the order of $T$, setting $q\sim T/\hbar s$ in \Eq{e8} results in a correct order-of-magnitude estimate of the typical scattering rate, see \Eq{e16} below.

Independently of the initial state, at $t\to\infty$ the distribution function $N_q$ relaxes to $n_q$. In order to study the approach to equilibrium, we substitute 
\beq
N_q = n_q+g_q f_q,
\quad
g_q = \sqrt{n_q(1+n_q)\,} 
\label{e9}
\eeq
into Eqs.~\eq{e6}--\eq{e7}, neglect all but linear in $f_q$ contributions, and obtain
\beqa
\frac{\partial f_q}{\partial t}
&=& \!-\,\frac{2\pi}{\hbar^2}\sum_p\!\sum_{q_1>q_2}\bigl|t_{qp;q_1q_2}\bigr|^2\!
\left(\frac{f_q}{g_q} + \frac{f_p}{g_p} - \frac{f_{q_1}}{g_{q_1}} - \frac{f_{q_2}}{g_{q_2}}\right)
\nonumber\\
&\times&\!\! g_p g_{q_1} g_{q_2} 
\delta(\omega_{q}+\omega_{p}-\omega_{q_1}-\omega_{q_2})\,\delta_{q+p,q_1+q_2}.
\label{e10} 
\\
\nonumber
\eeqa

The linearized Boltzmann equation \eq{e10} is applicable for both positive and negative $q$. Focusing from now on on $q>0$, we note that \Eq{e10} simplifies considerably if 
\beq
\xi(T/\hbar s)^3 \ll q \ll (T/\hbar s\xi)^{1/3}.
\label{e11}
\eeq  
The first inequality in \Eq{e11} ensures that contributions from the processes with all bosons but $q$ on the left-moving branch of the spectrum are exponentially suppressed. The second inequality in \Eq{e11} guarantees that the wave number of the only left-moving boson participating in the remaining scattering processes is much smaller than $T/\hbar s$. Under these conditions, the spectrum in the right-hand side of \Eq{e10} can be linearized, which amounts to neglecting corrections of order  $\xi(T/\hbar s)^2 \ll 1$ [this inequality is implicit in \Eq{e11}]. This approximation corresponds to the substitution into \Eq{e10}
\begin{widetext}
\beq
\delta\bigl(\omega_{q}+\omega_{p}-\omega_{q_1}-\omega_{q_2}\bigr)\,\delta_{q + p, q_1 + q_2}
\approx \frac{1}{2s}
\bigl[
\delta(p + 0)\,\delta_{q, q_1 + q_2} 
+ 
\delta(q_2 + 0)\,\delta_{q+p, q_1}
\bigr],
\label{e12}
\eeq
where $\delta(k+0)$ indicates that $k$ is an infinitesimal wave number on the left-moving branch. This yields
\beqa
\frac{\partial f_q}{\partial t}
&=& 
-\,\frac{1}{4\pi^3}\lambda^2 K^2 s \,(T/\hbar s)\,q\!\int_0^\infty \!\! dq_1
\,\Biggl\lbrace
\frac{1}{2}\!\int_0^\infty \!\! dq_2\,
\delta(q-q_1-q_2)\,
q_1g_{q_1} q_2  g_{q_2}\!
\left(\frac{f_q}{g_q}  - \frac{f_{q_1}}{g_{q_1}} - \frac{f_{q_2}}{g_{q_2}}\right)
\nn \\
&&
+ \,\int_0^\infty \!\! dp\,
\delta(q+p-q_1)\,
p_{} g_p q_1 g_{q_1} \!
\left(\frac{f_q}{g_q}  + \frac{f_{p}}{g_{p}} - \frac{f_{q_1}}{g_{q_1}}\right)
\Biggr\rbrace,
\label{e13}
\eeqa
\end{widetext}
where $g_q^\pprime, g_p^\pprime, g_{q_1}$, and $g_{q_2}$ are given by \Eq{e9} with a linearized spectrum, e.g., $g_q = \bigl[2\sinh(\hbar s q/2T)\bigr]^{-1}$. 
The factor $T/\hbar s$ in the right-hand side of \Eq{e13} is a remnant of the left-moving boson. Indeed, its wave number $k$ [$k$ is either $p$ or $q_2$, see \Eq{e12}] appears in \Eq{e10} in combination $|k| g_k$, where the factor $|k|$ comes from the square of the amplitude \eq{e5}. For $|k|\ll T/\hbar s$, we have $g_k = (T/\hbar s)|k|^{-1}$, which gives $|k| g_k = T/\hbar s$.
  
Note that the parameter $\xi$ [see \Eq{e3}] does not appear explicitly in \Eq{e13}. This is consistent with the above result for the relaxation rate of high-energy bosons: $\tau_q^{-1}$ is independent of $\xi$ at $q\ll (T/\hbar s\xi)^3$, see \Eq{e8}.
Note also that all wave numbers in \Eq{e13} are strictly positive: coupling between bosons moving in opposite directions appears only in higher orders in $\xi(T/\hbar s)^2$.
Accordingly, the right-hand side of \Eq{e13} involves only three bosons moving in the same direction. This kind of scattering processes has a divergent rate when the spectrum is taken to be strictly linear from the outset, see \Eq{1} and \Fig{Fig1}(a). While \Eq{e13} also corresponds to the limit of vanishing spectrum nonlinearity, it is crucial that the spectrum is linearized \textit{after} the scattering amplitudes are evaluated and the divergences are regularized. 

After integration over $q_2$ and $p$, \Eq{e13} assumes the form  
\beq
\frac{\partial}{\partial t}f(x,t)
= -\,\tau_0^{-1}\!\int_0^\infty\!dy\,\mathcal G(x,y) f(y,t),
\label{e14}
\eeq
where $f(x,t) = f_q(t)\bigr|_{q = 2\pi (T/\hbar s) x}$. The kernel $\mathcal G(x,y)$ is given by
\beqa
\mathcal G(x,y) &=& 
\frac{xy(x+y)}{\sinh\bigl[\pi(x+y)\bigr]}
- \frac{xy(x-y)}{\sinh\bigl[\pi(x-y)\bigr]}
\nn\\
&&+\, \frac{1}{6}x^2(x^2 + 1)\delta(x-y)\,,
\label{e15}
\eeqa
and the typical scattering rate is
\beq
\tau_0^{-1} 
=  2\pi\lambda^2 K^2 s \,(T/\hbar s)^5. 
\label{e16}
\eeq
The integro-differential equation \eq{e14}-\eq{e15} can be solved exactly. The solution reads~[\onlinecite{suppl}]
\beq
f(x,t) = \alpha_0\varphi_0(x) 
+ \!\int_0^\infty\!\!d\nu\,\alpha_\nu\varphi_\nu(x)\,e^{-\eta_\nu t/\tau_0},
\label{e17}
\eeq
where $\eta_\nu = \nu^2(\nu^2 + 1)/6$ and 
\beqa
\varphi_0(x) &=&  \sqrt{6\pi\,}\frac{x}{\sinh(\pi x)},
\label{e18}
\\
\varphi_\nu(x)&=& \frac{1}{\sqrt{(\nu^2 + 1)(4\nu^2 +1)}}
\biggl\lbrace 
(2\nu^2-1 )\delta(x-\nu)
\qquad~~
\nn\\
\nn\\
&&
+ \,\frac{3x}{\sinh[\pi(x+\nu)]}
+ \frac{3x}{\sinh[\pi(x- \nu)]}
\biggr\rbrace.
\label{e19}
\eeqa
(The singularity in the right-hand side of \Eq{e19} is to be understood as the principal value.)
The coefficients $\alpha_0$ and $\alpha_\nu$ in \Eq{e17} are determined  by the initial conditions, $\alpha_{\mu} = \int_0^\infty\!dx\,\varphi_\mu(x) f(x,0)$ for $\mu = 0,\nu$.

The first term in the right-hand side of \Eq{e17} represents a stationary (independent of $t$) contribution to $f(x,t)$. At $t\to\infty$ Eqs. \eq{e9} and \eq{e17} yield 
\beq
\delta N_q = N_q\bigr|_{t\to\infty} - n_q 
= \alpha_0 g_q\varphi_0(x)\bigr|_{x = \hbar s q/2\pi T}.
\label{e20}
\eeq 
This result has a clear physical meaning. In general, a stationary (equilibrium) solution of the Boltzmann equation $N_q\bigr|_{t\to\infty}$ is not unique. All such solutions, however, have the form of the Bose function $n_q$, parametrized by temperature $T$. A change of $T$ by $\delta T$ generates a correction to $N_q\bigr|_{t\to\infty}$, which, to linear order in $\delta T$, indeed has the form \eq{e20} with $\alpha_0 = \sqrt{\pi/6\,}(\delta T/T)$. On the other hand, the energy of the system at $t\to\infty$ coincides with that in the initial non-equilibrium state. Thus, the temperature $T$ characterizing the equilibrium distribution at $t\to\infty$ is uniquely determined by the initial conditions. Choosing $n_q$ as the Bose distribution with this temperature, one ensures that $\alpha_0 = 0$ in \Eq{e17}. 

The remaining (time-dependent) term in the right-hand side of \Eq{e17} describes approach to equilibrium. At short times, $t\ll\tau_0$, only the relaxation modes with $\nu\gtrsim (\tau_0/t)^{1/4}\gg 1$ are affected. Since $\varphi_\nu(x)\approx \delta(x-\nu)$ at $\nu\gg 1$, \Eq{e17} gives $f(x,t)\propto e^{-\eta_x t/\tau_0}$, which describes exponential relaxation with the rate given by the appropriate limit of \Eq{e8} [$q\ll (T/\hbar s \xi)^{1/3}$, see \Eq{e11}]. 

At $t\gg\tau_0$ the high-energy bosons have already relaxed, and thermal bosons (with $x\sim 1$ or $q\sim T/\hbar s$) have equilibrated among themselves, although at temperature that has not yet reached its equilibrium value. Indeed, at large $t$ the main contribution to the integral in \Eq{e17} comes from small $\nu$. Approximating $\varphi_\nu(x) \approx -\,\delta(x-\nu) + \sqrt{6/\pi\,}\varphi_0(x)$ and $\eta_\nu \approx \nu^2/6$, 
we find $\alpha_\nu = -f(\nu,0)$, and \Eq{e17} yields
\beq
f(x,t) = F(x,t) -  \sqrt{6/\pi\,}\varphi_0(x)\!\int_0^\infty\!\!d\nu\,F(\nu,t),
\label{e21}
\eeq
where $F(x,t) = f(x,0)\,e^{-x^2 t/6\tau_0}$ corresponds to exponential relaxation with the rate
\beq
\tau_q^{-1} = \frac{1}{12\pi}\lambda^2 K^2 s\,(T/\hbar s)^3 q^2.
\label{e22}
\eeq
The role of the second term in \Eq{e21} is to ensure the energy conservation. The corresponding correction to the distribution function [see \Eq{e9}] can be cast in the form 
\[
\delta N_q = \frac{\,\partial n_q}{\partial T}\,\delta T(t),
\quad
\delta T(t) = -\,\frac{3\hbar s\,}{\,\pi^2}\!\int_0^\infty\!\!dq\,f_q(0) \,e^{-t/\tau_q},
\]
with $1/\tau_q$ given by \Eq{e22}. For generic $f_q(0)$, the correction to temperature $\delta T(t)$ exhibits non-exponential dependence on time. 

To summarize, elementary excitations of one-dimensional interacting systems are often described in the framework of the effective Luttinger liquid theory. Both the conventional Luttinger liquid theory~[\onlinecite{Haldane,LL_bosons,Giamarchi}] and its recent extensions~[\onlinecite{PKKG,Pereira,IG}] provide a set of efficient tools for evaluation of various correlation functions. However, none of these approaches is capable of describing the thermalization of bosonic quasiparticles because interaction between bosons with linear spectrum results in a divergent inelastic scattering rate.

In this Letter we demonstrated that the divergences are regularized when the nonlinearity of boson spectrum is taken into account. We derived and solved the Boltzmann equation describing the fastest equilibration process in the system, namely, thermalization of bosons moving in the same direction. The equation describes bosons with a linearized (as opposed to strictly linear) spectrum and results in a finite relaxation rate that scales with temperature as $T^5$, see \Eq{e16}. Our results are applicable to both fermions and bosons with strong long-range repulsion.

The work at Argonne National Laboratory is supported by the U.S. DOE, Office of Science, under Contract No. DE-AC02-06CH11357.
K. A. M. and M. P. are grateful to the Aspen Center for Physics (NSF Grant No. 1066293), where part of this work was performed. M. P. thanks the Galileo Galilei Institute for Theoretical Physics and the INFN and the Kavli Institute for Theoretical Physics at UCSB (NSF Grant No. PHY11-25915) for their warm hospitality and partial support during the completion of this project. 



\onecolumngrid
\clearpage
\begin{center}
{\large\textbf{Thermalization of acoustic excitations \\ in a strongly interacting one-dimensional quantum liquid
\\~\\
Supplemental Material
}}

\setcounter{page}{1}
\thispagestyle{empty}

Jie Lin,$^1$ K. A. Matveev,$^1$ and M. Pustilnik$^2$

\small\textit{
$^1$Materials Science Division, Argonne National Laboratory, Argonne, IL 60439
\\
$^2$School of Physics, Georgia Institute of Technology, Atlanta, GA 30332
}
\end{center}
\twocolumngrid

In Sec.~\ref{phonons} of this Supplemental Material we derive the dispersion relation for phonons in the Wigner crystal. The derivation of the on-shell amplitude of the leading inelastic scattering process is discussed in detail in Sec.~\ref{amplitude}.  In particular, in Sec.~\ref{Coulomb} we address the screened Coulomb interaction, and in Sec.~\ref{integrable} we demonstrate that the amplitude vanishes for integrable models with $V(x)\propto 1/\sinh^2(c\rho x)$ and $V(x)\propto 1/x^2$. Finally, in Sec.~\ref{Boltzmann} we present the exact solution of the linearized Boltzmann equation.

\section{Phonons in a Wigner crystal}
\label{phonons}

\setcounter{equation}{0}
\numberwithin{equation}{section}

We consider a system of $N$ identical spinless particles of mass $m$ described by the Hamiltonian [see Eq.~(4) of the Letter] 
\beq
H=\sum_l\frac{p_l^2}{2m}+\frac{1}{2}\sum_{l,l^\prime} V(x_l-x_{l^\prime}),
\label{1.1}
\eeq
where $p_l$ and $x_l$ are, respectively, the momentum and position of the $l$th particle $(l = 1,\ldots,N)$, and $V(x)$ is the interaction potential. Expanding the potential energy in \Eq{1.1} to leading order in $|u_l - u_{l'}|$, we obtain the Hamiltonian of a harmonic chain, 
\beq
H_0=\sum_{l}\frac{p_l^2}{2m} +
\frac{1}{4}\sum_{l,l^\prime} V_{l-l^\prime}^{(2)}(u_l-u_{l^\prime})^2,
\label{1.2}
\eeq  
where we introduced the notation 
\beq
V_l^{(m)}=\left.\frac{d^mV(x)}{dx^m}\right |_{x=l/\rho}.
\label{1.3}
\eeq
It is convenient to write $u_l$ and $p_l$ in the second-quantized representation,
\beqa
u_l&=&\sum_{q} \sqrt{\frac{\hbar}{2mN\omega_q}}\,(b_q^\pdag+b_{-q}^\dagger)e^{iql},
\label{1.4}\\
p_l&=&-i\sum_{q} \sqrt{\frac{\hbar m\omega_q}{2N}}\,(b_q-b_{-q}^\dagger)e^{iql},
\label{1.5}
\eeqa 
where the phonon creation and annihilation operators satisfy the canonical commutation relation $\bigl[ b_q^\pdag,b_{q'}^\dagger\bigr] = \delta_{qq'}$. Substitution into \Eq{1.2} then yields
\beq
H_0=\sum_q\hbar\omega_q\bigl(b_q^\dagger b_q^\pdag +1/2\bigr),
\label{1.6}
\eeq
where the phonon frequencies $\omega_q$ are given by
\beq
\omega_q^2=\frac{2}{m} \sum_{l=1}^\infty V_l^{(2)}
\bigl[1-\cos(ql)\bigr].
\label{1.7}
\eeq
Since \Eq{1.7} does not contain $\hbar$, the same result can be obtained by solving classical equations of motion. 

At small $q$ and for $V(x)$ decaying with the distance faster than $1/x^3$, \Eq{1.7} reduces to Eq.~(3) of the Letter,
\beq
\omega_q = s|q|\bigl(1 - \xi q^2\bigr)
\label{1.8}
\eeq 
with 
\beq
s = \sqrt{\frac{V_{22}}{m}}\,,
\quad
\xi = \frac{1}{24}\frac{V_{24}}{V_{22}}\,,
\label{1.9}
\eeq
where 
\beq
V_{mn} = \sum_{l=1}^{\infty}V_l^{(m)}l^n.
\label{1.10}
\eeq
Modifications of \Eq{1.8} for potentials decaying with the distance as $1/x^3$ (screened Coulomb potential) and $1/x^2$ (Calogero-Sutherland model) are discussed below in Sec. \ref{Coulomb} and Sec. \ref{integrable}, respectively.

\section{Scattering amplitude}
\label{amplitude}

We start with the general expression for the on-shell scattering amplitude in order $\hbar^2$ derived in Ref.~[\onlinecite{MAP-s}],
\beq
t_{q_1^\pprime q_2^\pprime; q_1^\prime q_2^\prime}=\frac{\hbar^2}{m^3 N}\frac{\Lambda}{(\omega_{q_1}\omega_{q_2}\omega_{q_1^\prime}\omega_{q_2^\prime})^{1/2}}\,,
\label{B1}
\eeq
where 
\begin{widetext}
\beq
\Lambda = -\,\frac{f_3(q_1,q_2)f_3(q_1^\prime,q_2^\prime)}{\omega_{q_1+q_2}^2-(\omega_{q_1}+\omega_{q_2})^2}+\frac{f_3(q_2,-q_1^\prime)f_3(q_1,-q_2^\prime)}{\omega_{q_2-q_1^\prime}^2-(\omega_{q_2}-\omega_{q_1^\prime})^2}
+\frac{f_3(q_1,-q_1^\prime)f_3(q_2,-q_2^\prime)}{\omega_{q_2-q_2^\prime}^2-(\omega_{q_2}-\omega_{q_2^\prime})^2}+\frac{m}{2}f_4(q_1,q_2,-q_1^\prime)
\label{B2}
\eeq
with
\beqa
f_3(q_1,q_2) &=& 
\sum_{l=1}^\infty V_l^{(3)}\Bigl\{\sin[(q_1+q_2)l]
-\sin(q_1l) -\sin(q_2l)\Bigr\},
\label{B3}
\\
f_4(q_1,q_2,q_3) 
&=& \sum_{l=1}^\infty V_l^{(4)}\Bigl\{1-\cos(q_1l)-\cos(q_2l)
-\cos(q_3l)-\cos[(q_1+q_2+q_3)l]
\label{B4} \\
&& \qquad\qquad\qquad\qquad
+\cos[(q_1+q_2)l]+\cos[(q_2+q_3)l]
+\cos[(q_1+q_3)l]\Bigr\}.
\nonumber
\eeqa
Expanding Eqs. \eq{B2}--\eq{B4} to first order in $q_1^\pprime,q_2^\pprime,q_1^\prime,q_2^\prime$, and taking into account the momentum and energy conservation laws, 
$q_1 + q_2 = q_1^\prime + q_2^\prime$ and
$\omega_{q_1} + \omega_{q_2} = \omega_{q_1^\prime} + \omega_{q_2^\prime}$,
we arrive at Eq.~(5) of the Letter,
\beq
t_{q_1^\pprime q_2^\pprime;q_1^\prime q_2^\prime}
= \frac{\lambda}{N} \frac{\hbar^2 \rho^2}{m\,}\bigl|q_1^\pprime q_2^\pprime q_1^\prime q_2^\prime\bigr|^{1/2}
\label{A4}
\eeq
with
\beq
\lambda =
\frac{V_{22}V_{44}-V_{33}^2}{4\rho^2 V_{22}^2} 
+ \frac{V_{33}^2}{16\rho^2 V_{22}^2}
\lim_{q,q' \to 0}\left\{
\frac{A^\prime_{q+q'}-A^\prime_q-A^\prime_{q'}}{\mathcal F[A_q]}
-\,\frac{4\rho V_{22}}{V_{33}}\frac{\mathcal F[B_q]}{\mathcal F[A_q]}
\right\}.
\label{A5}
\eeq
Here
\beq
A_q=\omega_q^2,
\quad
A^\prime_q = \frac{dA_q}{dq}\,,
\quad
B_q = \frac{2}{m\rho}\sum_{l=1}^\infty V_l^{(3)}l\bigl[1-\cos(ql)\bigr],
\label{A6}
\eeq
\end{widetext}
and the functional $\mathcal F$ is defined as
\beq
\mathcal F[f(q)] = \frac{f(q+q')}{q+q'} - \frac{f(q)}{q}- \frac{f(q')}{q'}.
\eeq

Interestingly, although the right-hand side of \Eq{A5} depends on the functional form of the interaction potential, it is independent of the interaction strength: multiplication of $V(x)$ by an arbitrary constant leaves $\lambda$ invariant.

Equation \eq{A5} is valid provided that $V_{nn}$ are finite, i.e., for interaction potentials decaying with the distance faster than $1/x$. Further simplification is possible if $V_{n,n+2}$ [see \Eq{1.10}] are finite as well, i.e., for $\lim_{x \to\infty} x^3 V(x) = 0$. Expanding $A(q)$ and $B(q)$ in \Eq{A5} to fourth order in $q$, we find
\beq
A_q = s^2 q^2(1 - 2\xi q^2),
\quad
B_q = 
\frac{V_{33}q^2}{m\rho}\left(1-\frac{V_{35}q^2}{12V_{33}}\right)
\label{A7}
\eeq
with $s$ and $\xi$ given by \Eq{1.9}. Substitution of \Eq{A7} into \Eq{A5} results in 
\beq
\lambda = \frac{V_{24}V_{44} - V_{33}V_{35}}{4\rho^2 V_{22} V_{24}}\,.
\label{A88}
\eeq

\subsection{Screened Coulomb interaction}
\label{Coulomb}

We turn now to the important in practice case of the Coulomb potential screened by a remote gate at distance $d$ from the Wigner crystal,
\beq
V(x) = \frac{e^2}{|x|} - \frac{e^2}{\sqrt{x^2 + 4d^2\,}}
\label{A16}
\eeq
For this potential $V_{nn}$ are finite and determined by the behavior of the potential at short distances $x\lesssim d$. For $\rho d\gg 1$ we find
\beq
V_{nn} = (-1)^n n! \,e^2\!\rho^{n+1}\ln(\rho d).
\label{A17}
\eeq
$V_{n,n+2}$, however, diverge due to slow decay of $V(x)$ at $x\gg d$,
\beq
V_{n,n+2} = (-1)^n (n+2)! \,e^2 d^2 \rho^{n+3}S, 
\label{A18}
\eeq 
where $S = \sum_{l = \rho d}^{\infty} l^{-1}$ is a logarithmically divergent sum. (In writing \Eq{A18}, we neglected the contribution from $l <\rho d$.) Treating $S$ as a large, but finite number (or, equivalently, introducing a cutoff $l<l_0$ with $l_0\gg\rho d$), we find $V_{35}/V_{24} = -5\rho$. Substituting this relation into \Eq{A88} and taking into account \Eq{A17}, we obtain
\beq
\lambda = - 3/4.
\label{A19}
\eeq

An accurate derivation of \Eq{A19} should rely on \Eq{A5} rather than \Eq{A88}.
Even though $V_{24}$ diverges, the dispersion relation can still be cast in the form of \Eq{1.7}, but with $q$-dependent $\xi$. For small $q$, 
\[
q\ll 1/\rho d\ll 1,
\] 
we find
\beq
\omega_q = s q\bigl(1 - \xi_q q^2\bigr)
\label{A20}
\eeq
with
\beq
s = \sqrt{(2e^2\!\rho^3\!/m)\ln(\rho d)\,}\,,
\quad
\xi_q = \frac{(\rho d)^2}{2\ln(\rho d)}\ln(1/q).
\label{A21}
\eeq
In the same limit, the functions $A_q$ and $B_q$, see \Eq{A6}, are given by 
\beq
A_q = s^2\!q^2\bigl(1 - 2\xi_q q^2\bigr),
\quad
B_q = - s^2\!q^2\bigl(3 - 5\xi_q q^2\bigr).
\label{A22}
\eeq
Substituting \Eq{A22} into \Eq{A5} and taking into account \Eq{A17}, we recover \Eq{A19}.

\subsection{Integrable models}
\label{integrable}

In integrable models~[\onlinecite{Sutherland-s}] conservation laws forbid a redistribution of momenta between colliding particles. In other words, collisions do not lead to relaxation in such models. This lack of relaxation should not depend on whether the system is described in terms of the bare particles or in terms of the quasiparticles. In accordance with this observation, the amplitude \eq{B1}, as well as the amplitudes of higher-order scattering processes, are expected to vanish for all integrable models. This vanishing of the scattering amplitude also serves as an independent check of the validity of Eqs. \eq{A5} and \eq{A88} above. 
 
Consider the potential~[\onlinecite{Sutherland-s}] 
\beq
V(x)=\frac{V_0}{\sinh^2(c\rho x)}\,.
\label{A8}
\eeq
For this potential 
\[
\frac{\partial^m}{\partial x^m}V(x) = c^m \frac{\partial^m}{\partial c^m}V(x)\, x^{-m},
\]
and Eqs. \eq{1.3} and \eq{1.10} yield
\beq
V_{mn} = (c\rho)^m \frac{d^m}{dc^m} V_{0, n-m}\,.
\label{A91}
\eeq
Another useful relation, 
\beq
V_{02} = -\,\frac{1}{2}\frac{d}{dc}V_{00}\,,
\label{A92}
\eeq
can be derived by comparing the expansions
\[
V_{0n} = 4V_0 \sum_{l=1}^\infty\sum_{m=1}^\infty l^n m \,e^{-2clm}
\] 
for $n=0$ and $n=2$.
Using the identities \eq{A91} and \eq{A92}, we find
\beq
\frac{V_{nn}}{\,(c\rho)^n} = - \,2\frac{V_{n-1,n+1}}{(c\rho)^{n-1}} 
= \frac{\partial^n V_{00}}{\partial c^n},
\eeq
which gives
\beq
V_{24}V_{44} = V_{33}V_{35} 
= - \,\frac{\,(c\rho)^6}{2} 
\frac{\partial^3 V_{00}}{\partial c^3}\frac{\partial^4 V_{00}}{\partial c^4}.
\eeq
Substitution into \Eq{A88} then yields $\lambda = 0$.

At $c\gg 1$ and in the Wigner crystal limit $V_0\gg\hbar^2\!\rho^2 e^{2c}/mc^2$, the potential \eq{A8} realizes the \textit{Toda lattice} model. In this regime 
\beq
V(x) = 4V_0 e^{-2c\rho|x|}
\eeq
and all but $l=1$ contributions to the sum in \Eq{1.10} can be neglected (this corresponds to the nearest neighbors interaction in the Wigner crystal). With these approximations,
\beq
V_{mn} = 4V_0(-1)^m(2c\rho)^m e^{-2c},
\eeq
and the relation $V_{24}V_{44} = V_{33}V_{35}$ is obvious.

At $c\ll 1$ the potential \eq{A8} becomes
\beq
V(x) = \frac{\alpha\,}{\,x^2}
\label{A13}
\eeq
with $\alpha = V_0(c\rho)^{-2}$ ($\alpha\gg\hbar^2/m$ in the Wigner crystal regime),
which corresponds to the integrable \textit{Calogero-Sutherland} model~[\onlinecite{Sutherland-s}]. 
For this potential $V_{nn}$ are finite, 
\beq
V_{nn} = (-1)^n (n+1)! (\pi^2/6)\alpha\rho^{n+2},
\label{A14}
\eeq
but $V_{n,n+2}$ diverge, and \Eq{A88} is inapplicable. 

Equation \eq{A8} can be viewed as a version of \Eq{A13} with a long-distance cutoff. Importantly, for the Calogero-Sutherland model the sum entering $V_{n,n+2}$ [see \Eq{1.10}] diverges as a power-law rather than logarithmically, as is the case for the screened Coulomb interaction, see Sec.~\ref{Coulomb}. Accordingly, the value of $\lambda$ depends on the cutoff scheme and is not universal. It is therefore important to demonstrate vanishing of the scattering amplitude directly for the Calogero-Sutherland model \eq{A13} instead of relying on $c\to 0$ limit of \Eq{A8}. 

For the potential \eq{A13} we find 
\beq
A_q = -\, \frac{1}{4}\, B_q
= s^2q^2\!\left(1-\frac{q}{2\pi}\right)^2,
\quad 
s^2 = \frac{\alpha\pi^2 \! \rho^4}{m}.
\label{A15}
\eeq
Substituting \Eq{A15} into \Eq{A5} and using \Eq{A14}, we indeed find $\lambda = 0$.
Note that for the Calogero-Sutherland model
\beq
\omega_q = \sqrt{A_q} = sq\,(1-q/2\pi)^2,
\eeq
i.e., the nonlinear term in the dispersion relation is quadratic in $q$ rather than cubic as in \Eq{1.8}; this property is specific~[\onlinecite{Sutherland-s}] for the inverse-square interaction potential. 


\section{Solution of the linearized Boltzmann equation}
\label{Boltzmann}

Solutions of the equation [see Eqs.~(14)--(15) of the Letter]
\beqa
\frac{\partial}{\partial t}f(x,t)
&=& -\,\tau_0^{-1}\!\int_0^\infty\!dy\,\mathcal G(x,y) f(y,t).
\label{5.1}
\\
\mathcal G(x,y) &=& 
\frac{xy(x+y)}{\sinh\bigl[\pi(x+y)\bigr]}
- \frac{xy(x-y)}{\sinh\bigl[\pi(x-y)\bigr]}
\nn\\
&& \quad+\,\frac{1}{6}x^2(x^2 + 1)\delta(x-y) 
\label{5.2}
\eeqa
have the form $f(x,t) = \varphi (x)\exp(-\eta t/\tau_0)$,
where $\varphi(x)$ satisfies the integral equation
\beq
\eta\varphi(x) = \int_0^\infty\!dy\,\mathcal G(x,y)\varphi(y).
\label{5.3}
\eeq  

Remarkably, the eigenvalue problem \eq{5.3} can be solved exactly. First, we formally extend $\varphi(x)$ to negative $x$ according to 
\beq
\varphi(-x) = \varphi(x)
\label{5.4}
\eeq
and rewrite \Eq{5.3} as
\beq
\eta\varphi(x) = \frac{1}{6}x^2(x^2 + 1) \varphi(x)
+ \int_{-\infty}^\infty\!\!dy\, 
\frac{xy(x+y)}{\sinh\bigl[\pi(x+y)\bigr]}\,\varphi(y),
\label{5.5}
\eeq
Next, we multiply both sides of \Eq{5.5} by $e^{i\zeta x}$ and integrate over $x$. This transforms the integral equation \eq{5.5} into a differential equation, which can be written as
\beq
\bigl(h_\zeta^2 + h_\zeta\bigr)\widetilde\varphi(\zeta) = 6\eta\,\widetilde\varphi(\zeta).
\label{5.6}
\eeq
Here $\widetilde\varphi(\zeta)$ is the Fourier transform of $\varphi(x)$,
\beq
\widetilde\varphi(\zeta) = \widetilde\varphi(- \zeta) =\int\!dx\,e^{i\zeta x}\varphi(x),
\label{5.7}
\eeq
and the operator $h_\zeta$ is given by
\beq
h_\zeta = -\,\frac{d^2}{d\zeta^2} - \frac{3}{2\cosh^2(\zeta/2)}\,.
\label{5.8}
\eeq

Equation \eq{5.8} coincides with the Hamiltonian of a particle moving in one dimension in the presence of the reflectionless P\"oschl-Teller potential. The corresponding eigenvalue problem,
\beq
h_\zeta\psi_\epsilon(\zeta) = \epsilon_{}\psi_\epsilon(\zeta),
\label{5.9}
\eeq 
is discussed in detail in, e.g.,~[\onlinecite{Schwabl}]. 
Even-parity eigenstates of $h_\zeta$, which are of interest here [see \Eq{5.7}], include the ground state with the eigenvalue $\epsilon = -1$,
\beq
\psi_{-1}(\zeta) = \frac{1}{\cosh^2(\zeta/2)},
\label{5.10}
\eeq
and a continuum of states with eigenvalues $\epsilon = \nu^2$,
\beq
\psi_{\nu^2} (\zeta) 
= \left(-\,\frac{d}{d\zeta} + \tanh\frac{\zeta}{2}\right)
\!\!\left(-\,\frac{d}{d\zeta} + \frac{1}{2}\tanh\frac{\zeta}{2}\right)
\cos(\nu\zeta)
\label{5.11}
\eeq

Obviously, the eigenstates of $h_\zeta$, see \Eq{5.9}, are also eigenstates of $h_\zeta^2 + h_\zeta$, see \Eq{5.6}. The corresponding eigenvalues are related according to 
$\eta = \epsilon(\epsilon +1)/6$, which gives $\eta_0 = 0$ for the bound state ($\epsilon = -1$), and $\eta_\nu = \nu^2(\nu^2+1)/6$ for the continuum ($\epsilon = \nu^2$). 

Carrying out the inverse Fourier transform of Eqs. \eq{5.10} and \eq{5.11}, we find Eqs.~(18) and (19) of the Letter,
\beqa
\varphi_0(x) &=& \sqrt{6\pi\,}\frac{x}{\sinh(\pi x)},
\label{5.12}
\\
\varphi_\nu(x)&=& \frac{1}{\sqrt{(\nu^2 + 1)(4\nu^2 +1)}}
\biggl\lbrace 
(2\nu^2-1 )\delta(x-\nu)
\qquad\quad
\nn\\
\nn\\
&&\quad
+ \,\frac{3x}{\sinh[\pi(x+\nu)]}
+ \frac{3x}{\sinh[\pi(x- \nu)]}
\biggr\rbrace,
\label{5.13}
\eeqa
with the singularity in the right-hand side of \Eq{5.13} understood as the principal value. 

The eigenfunctions \eq{5.12} and \eq{5.13} are normalized according to 
\beq
\int_0^\infty\!dx\,\varphi_0^2(x) = 1,
\quad
\int_0^\infty\!dx\,\varphi_\nu(x)\varphi_{\nu'}(x) 
= \delta(\nu - \nu')
\label{5.14}
\eeq
[$\varphi_0(x)$ is orthogonal to $\varphi_\nu(x)$ for any $\nu$],
and form a complete set,
\beq
\varphi_0(x)\varphi_0(y) + \int_0^\infty\!\!d\nu\,\varphi_\nu(x)\varphi_\nu(y) = \delta(x-y).
\label{5.15}
\eeq
Thus, the general solution of Eqs. \eq{5.1}--\eq{5.2} can be written as an expansion in $\varphi_0$ and $\varphi_\nu$,
\beq
f(x,t) = \alpha_0\varphi_0(x) 
+ \int_0^\infty\!d\nu\,\alpha_\nu\varphi_\nu(x) e^{-\eta_\nu t/\tau_0},
\label{5.16}
\eeq
see Eq.~(17) of the Letter.
 


\begin{thebibliography}{50}

\bibitem{Giamarchi} 
T. Giamarchi, \textit{Quantum Physics in One Dimension} (Clarendon Press, Oxford, 2004).

\bibitem{Pines} 
D. Pines and P. Nozi\'eres, \textit{The Theory of Quantum Liquids} (Perseus Books, Reading, MA, 1994). 

\bibitem{Haldane} 
F. D. M. Haldane, 
Phys. Rev. Lett. \textbf{45}, 1358 (1980); 
J. Phys. C \textbf{14}, 2585 (1981).

\bibitem{LL_bosons}
F. D. M. Haldane, Phys. Rev. Lett. \textbf{47}, 1840 (1981);
V. N. Popov, Theor. Math. Phys. \textbf{11}, 565 (1972).

\bibitem{drag}
M. Pustilnik, E. G. Mishchenko, L. I. Glazman, and A. V. Andreev,
Phys. Rev. Lett. \textbf{91}, 126805 (2003).

\bibitem{PKKG}
M. Pustilnik, M. Khodas, A. Kamenev, and L. I. Glazman, 
Phys. Rev. Lett. \textbf{96}, 196405 (2006).

\bibitem{Pereira}
R. G. Pereira, J. Sirker, J.-S. Caux, R. Hagemans, J. M. Maillet, S. R. White, and I. Affleck,
Phys. Rev. Lett. \textbf{96}, 257202 (2006);
J. Stat. Mech. (2007) P08022.

\bibitem{mass}
For Galilean-invariant systems the effective mass $m_*$ satisfies~[\onlinecite{Pereira}]
$m/m_* = (4K)^{-1/2}(\rho/s)(\partial s/\partial\rho)$.
Since $\partial s/\partial\rho\sim s/\rho$ for $V(x)$ decaying with the distance as a power law, this yields $m_*/m\sim\sqrt{K}$.

\bibitem{Meyer-09} 
J. S. Meyer and K. A. Matveev, J. Phys.: Condens. Matter \textbf{21}, 023203 (2009).

\bibitem{Schulz-93} 
H. J. Schulz, Phys. Rev. Lett. \textbf{71}, 1864 (1993).

\bibitem{MAP} 
K. A. Matveev, A. V. Andreev, and M. Pustilnik, 
Phys. Rev. Lett. \textbf{105}, 046401 (2010);
Physica B \textbf{401}, 1898 (2012).

\bibitem{suppl}
See supplemental material for details.

\bibitem{Sutherland}
B. Sutherland, \textit{Beautiful Models}
(World Scientific, Singapore, 2004).

\bibitem{Lifshitz} 
E. M. Lifshitz and L. P. Pitaevskii, 
\textit{Physical Kinetics} (Pergamon Press, Oxford, 1981). 

\bibitem{IG}
A. Imambekov, T. L. Schmidt, and L. I. Glazman, arXiv:1110.1374.
  
\end{thebibliography}

\begin{thebibliography}{99}

\bibitem{MAP-s} 
K. A. Matveev, A. V. Andreev, and M. Pustilnik, 
Phys. Rev. Lett. \textbf{105}, 046401 (2010);
Physica B \textbf{401}, 1898 (2012).

\bibitem{Sutherland-s}
B. Sutherland, \textit{Beautiful Models}
(World Scientific, Singapore, 2004).

\bibitem{Schwabl}
F. Schwabl, \textit{Quantum Mechanics, 4th ed.} (Springer-Verlag, Berlin, Heidelberg, New York, 2007), Chapter 19.
 
\end{thebibliography}
\end{document}